\documentclass[%
reprint,
showpacs,
floatfix,
amsmath,amssymb,
]{revtex4-1}

\usepackage{color}
\usepackage{graphicx}
\usepackage{dcolumn}
\usepackage{bm}

\newcommand*{\citen}[1]{%
  \begingroup
    \romannumeral-`\x 
    \setcitestyle{numbers}%
    \cite{#1}%
  \endgroup   
}

\DeclareMathOperator{\sech}{sech}

\def\eeq{\relax}
\def\beq#1#2\eeq{\begin{equation}\label{#1}#2\end{equation}}
\def\bal#1#2\eal{\begin{align}\label{#1}#2\end{align}}
\def\bse#1#2\ese{\begin{subequations}\label{#1}#2\end{subequations}}

\begin{document}


\title{Broadband focusing of underwater sound using a transparent pentamode lens} 

\author{Xiaoshi Su}
 \email{xiaoshi.su@rutgers.edu}
 \affiliation{Mechanical and Aerospace Engineering, Rutgers University, Piscataway, NJ 08854}
\author{Andrew N. Norris}%
\affiliation{Mechanical and Aerospace Engineering, Rutgers University, Piscataway, NJ 08854}

\author{Colby W. Cushing}
\affiliation{Department of Mechanical Engineering and Applied Research Laboratories, The University of Texas
at Austin, Austin, Texas 78712}
\author{Michael R. Haberman}
\affiliation{Department of Mechanical Engineering and Applied Research Laboratories, The University of Texas
at Austin, Austin, Texas 78712}
\author{Preston S. Wilson}
\affiliation{Department of Mechanical Engineering and Applied Research Laboratories, The University of Texas
at Austin, Austin, Texas 78712}
\date{\today}

\begin{abstract}
We report an inhomogeneous acoustic metamaterial lens based on spatial variation of refractive index for broadband focusing of underwater sound. The index gradient follows a modified hyperbolic secant profile designed to reduce aberration and suppress side lobes. The gradient index (GRIN) lens is comprised of transversely isotropic hexagonal microstructures with tunable {\it quasi-static} bulk modulus and mass density. In addition, the unit cells are impedance-matched to water and have in-plane shear modulus  negligible compared to the effective bulk modulus. The flat GRIN lens is fabricated by cutting hexagonal centimeter scale hollow microstructures in aluminum plates, which are then stacked and sealed from the exterior water. Broadband focusing effects are observed within the homogenization regime of the lattice in both finite element (FEM) simulations and underwater measurements (20-40 kHz). This design approach has potential applications in medical ultrasound imaging and underwater acoustic communications.
\end{abstract}

\maketitle


\section{\label{intro}Introduction}

The quality of focused sound through a conventional Fresnel lens is usually limited by spherical/cylindrical aberration. Recent advances in acoustic metasurface design made it possible to manipulate the transmitted wavefront in an arbitrary way by achieving phase delay using space coiling structures. \cite{Li2012a,Xie2014,Li2014,Wang2014,Li2015} The aberration of the focused sound can be reduced by tuning the phase of the transmitted wave through simple ray tracing. However, this diffraction based design approach usually suffers from unbalanced impedance \cite{Estakhri2016} which is crucial to achieve destructive interference for canceling out side lobes. Therefore, this design approach requires more sophisticated modeling. \cite{Li2016} Many efforts have been made to achieve extraordinary transmission, \cite{Moleron2014,Tang2015} but the underlying physics is to tune the structure to achieve certain phase gradient of the transmitted wave at a particular frequency which limits the bandwidth of operation. Another disadvantage of the metasurface design is that the device only works at the steady state. \cite{Estakhri2016} In other words, it can not focus a pulse to a single focal spot.
Apart from the aforementioned disadvantages, the space coiling structure is not applicable for underwater devices because of the low contrast between bulk modulus of common materials and water. Both the fluid phase and the solid phase are connected to the background fluid, the existence of the Biot fast and slow compressional waves \cite{Biot1956,Biot1962} might cause strong aberration and induce more side lobes, while the shear mode will cause undesired scattering. Thus, we need to employee an alternative design method to overcome these issues.

The hyperbolic secant index profile has been widely used in GRIN lens designs. \cite{GRINOPTICS} \citet{Lin09} showed that the frequency independent analytical ray trajectories intersect at the same point, and demonstrated that it can be used in phononic crystal design to focus sound inside the device without aberration. \citet{Climente2010} adopted this approach in sonic crystal design, and experimentally demonstrated the broadband focusing effect beyond the lens with low aberration. Many other designs used the same index profile to focus airborne sound \cite{Zigoneanu2011,Romero-Garcia2013,Park2016} and underwater sound. \cite{Martin10} Most of the designs are based on variation of the filling fraction to achieve different refractive indices which usually cause significant impedance mismatch. Although transmission is not a big concern in many applications, it is determinant in the focusing capability of the GRIN lens. The focal distance is derived from ray tracing which is a transient solution. Nevertheless, the steady state focusing properties of the lens can be altered due to impedance mismatch between the lens and background medium. One exception is that \citet{Martin2015} modified the index distribution to reduce aberration and achieved high transmission by using hollow aluminum shells in a water matrix. However, the idea of adjusting the filling fraction introduces anisotropy and limits the range of effective properties which restrict the focal spot to be far from the lens.

In this paper, we utilize a two-dimensional (2D) version of the pentamode material (PM) \cite{Milton95,Norris11mw} to achieve a wide range of refractive indices, and introduce a new modification of the index profile for further aberration reduction. The advantage of PMs is that they can be designed to match the acoustic impedance to water and minimize the shear modulus which is undesired in acoustic designs, thus are very promising in underwater applications. For instance, \citet{Hladky-Hennion13} tuned the effective acoustic properties to water and experimentally demonstrated negative refraction at the second compressional mode. The structure is versatile such that it can be designed to achieve strong anisotropy, \cite{LaymanOrris2012} therefore is also a good choice for acoustic cloaking. \cite{Norris08b,Chen2015} In our design, the unit cells are transversely isotropic with index varying along the incidence plane. The modification of the index profile is done by using a one-dimensional coordinate transformation, the aberration reduction can be clearly observed from ray trajectories. The unit cells of the GRIN lens are designed using a static homogenization technique based on FEM \cite{Hassani98I} according to the modified index profile with a range from $0.5$ to $1$. Moreover, all the unit cells are impedance matched to water which is the key to obtain optimal focusing effect. The GRIN lens is fabricated by cutting centimeter scale hollow microstructures on aluminum plates using waterjet, then stacking and sealing them together. The interior of the compact solid matrix lens is filled with air, only the exterior faces are connected to water. The acoustic waves in the exterior water background are fully coupled to the structural waves inside the lens so that the lens is backscattering free and is capable of focusing sound as predicted. The GRIN lens is experimentally demonstrated to be capable of focusing underwater sound with high efficiency from $25$ kHz to $40$ kHz. The present design has potential applications in ultrasound imaging and underwater sensing where the water environment is important. The successful demonstration of our GRIN lens also shed light on the realization of pentamode acoustic cloak. \cite{Norris08b,Chen2015}

\section{\label{Gradient}Design of gradient index}

\subsection{Focal distance}

\begin{figure}[ht]
     \includegraphics[width=\columnwidth]{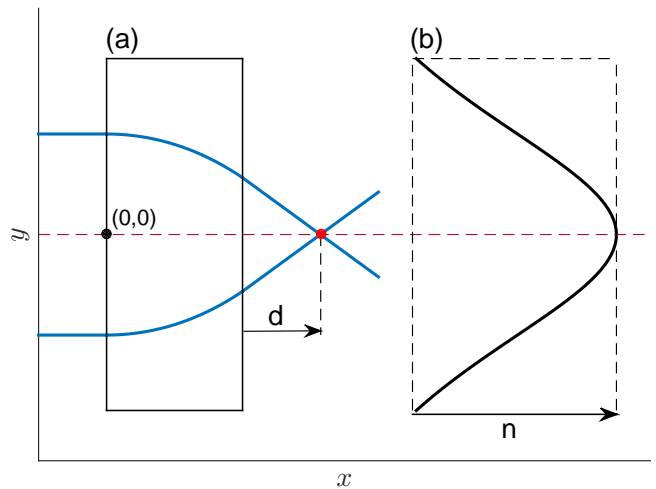}  
		\caption{A schematic view of the grin lens is shown in (a), along with two ray paths which focus a distance d from the lens surface.  The corresponding index of refraction profile within the lens is shown in (b).}\label{config}
\end{figure}

The rectangular outline of the 2D flat GRIN lens is designed as depicted in Fig.\ \ref{config} with  index profile symmetric with respect to the $x$-axis $(y=0)$. 
Assuming that the refractive index $n$ is a function only of $y$, the
 trajectories of a normally incident wave can be derived by solving a ray equation for $y = y(x)$ based on the fact that the component of slowness along the interface between each layer is constant:
\beq{0}
\frac{n \big(y(x)\big) } {\sqrt{1+y^{\prime 2}(x) }}
= n (y_0)  
\eeq
where $y_0 = y(0)$ is the incident position  on the $y$-axis at the left side of the lens, $x=0$. 
The focal distance from the right-hand boundary of the GRIN lens at $x=t$ is  
\begin{equation}\label{1}
d=y_t\, \sqrt{\frac{1}{n^2 (y_t) - n^2 (y_0)} -1} .
\end{equation}

\subsection{Hyperbolic secant and quadratic profiles}

We first consider a hyperbolic secant index profile $n(y)$:
\begin{equation}\label{profile}
n(y)=n_0\sech (\alpha y),
\end{equation}
where $n_0$  and $\alpha$ are constants. 
This profile, also  known as a Mikaelian lens, \cite{Mikaelian1980}  was originally proposed by  Mikaelian \cite{Mikaelian1951}  for both rectangular and cylindrical coordinates, and is often used  to design for low aberration. \cite{Lin09,Climente2010,Zigoneanu2011,Romero-Garcia2013,Park2016} 
 The ray trajectory is 
\begin{equation}\label{trajectory}
y(x)=\frac{1}{\alpha}\sinh^{-1}[\sinh(\alpha y_0)\cos(\alpha x)].
\end{equation}

Alternatively, consider the quadratic index profile \cite{Martin2015}
\begin{equation}\label{2.1}
n(y)=n_0\sqrt{1 - (\alpha y)^2 },
\end{equation}
for which the rays are 
\beq{2.2}
y(x) = y_0 \sqrt{2} \sin\Big( \frac{\pi}4 - \frac{n_0\, \alpha  x}{n(y_0)}\Big) .
\eeq

\citet{Martin2015} noted that the  above two profiles have opposite aberration tendencies,  and proposed a  mixed combination which shows reduced aberration.  However, in our design we are interested in a wider range in index, from unity to about $0.5$ (unlike Ref.~\citen{Martin2015} for which the minimum is $1/1.3 = 0.77$).  This requires $\alpha y_0$ to exceed unity, which rules out the use of the quadratic profile. It is notable that the purpose of using a wider range of index is to fully exploit the bulk space of the GRIN lens to achieve near field focusing capability.

\subsection{Reduced aberration  profile}
\begin{figure}[ht]
\centering
     \includegraphics[width=\columnwidth]{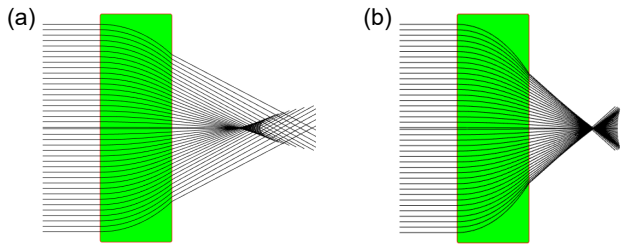}  
\caption{Ray tracing comparison between (a) the hyperbolic secant profile and (b) the reduced aberration profile.}\label{stretch}
\end{figure}

Here we use  a modified hyperbolic secant profile   by stretching the $y-$coordinate, as follows: 
\begin{equation}\label{2.4}
\begin{aligned}
n(y)& =n_0\sech \big( g(\alpha y)\big)
\ \ \text{where} \\
 g(z) & 
=  z / \big(1+\beta_1z^2 + \beta_2 z^4\big).
\end{aligned}
\end{equation}
The objective is to make $d$ of Eq.\ \eqref{1} independent of $y_0$ as far as possible. For small  $\alpha y_0$ we have from both Eqs.\ \eqref{trajectory} and \eqref{2.2} that $y(x) \approx  y_0 \cos \alpha x$, and hence for all three profiles 
\begin{equation}\label{3}
d \to d_0 \equiv   \frac 1{n_0\alpha} \cot \alpha t \ \ \text{as} \   \alpha y_0 \to 0 . 
\end{equation}
Note that $d_0 $ is independent of $y_0$, as expected.   This is the value of the focal distance that the modified profile \eqref{2.4} attempts to achieve for all values of  $y_0$ in the device by selecting  suitable values of the non-dimensional parameters $\beta_1$ and $\beta_2$.  Numerical experimentation led to the choice $\beta_1=-0.0679$ and $\beta_2=-0.002$. As a demonstration of aberration reduction, we plot the ray trajectories with and without the stretch in the $y-$direction are shown in Fig.\ \ref{stretch} for comparison. 
It is clear that the modified secant profile is capable of focusing a normally incident plane wave with minimal aberration.

\section{\label{Homo}Design of unit cells}
\begin{figure}[ht]
\centering
     \includegraphics[width=0.95\columnwidth]{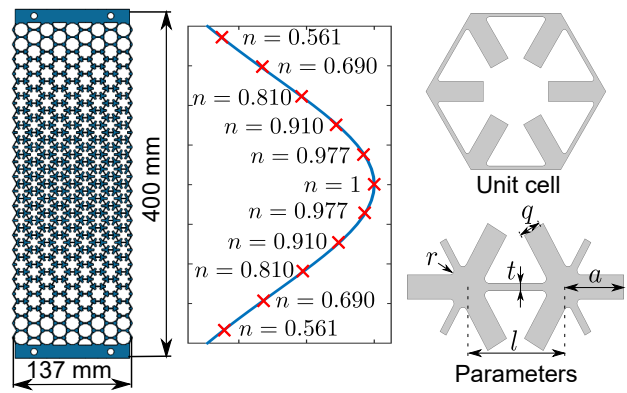} 
\caption{Pentamode lens design. The picture on the left side shows the top view of the designed lens, the plot in the middle shows the discretized index distribution within the lens, and the right side shows the unit cell structure and parameters.}\label{layered}
\end{figure}

The flat GRIN lens is designed using six types of unit cells corresponding to the discrete values selected from the modified hyperbolic index profile. Figure \ref{layered} shows the spatial distribution of refractive indices of the lens. The unit cell structure is the regular hexagonal lattice which has in-plane isotropy at the {\it quasi-static} regime. \cite{Norris2014} 
Using Voigt notation, the 2D pentamode elasticity requires $C_{11}C_{22}\approx C_{12}^2$ and $C_{66}\approx 0$ to minimize the shear modulus. With these requirements satisfied, the main goal is to tune the effective $C_{11}$ and mass density at the homogenization limit to achieve the required refractive index and match the impedance to water simultaneously. The material properties of water are taken as bulk modulus $\kappa_0=2.25$ GPa and density $\rho_0=1000$ kg/m$^3$. The material of the lens slab is aluminum with Young's modulus $E=70$ GPa, density $\rho=2700$ kg/m$^3$ and Poisson's ratio $\nu=0.33$. The geometric parameters of each unit cell, as shown in Fig.\ \ref{layered}, are predicted using foam mechanics \cite{KimHassani} and iterated using a homogenization technique based on FEM. \cite{Hassani98I} The geometric parameters of the six types of unit cells are listed in Table \ref{tab}. Note that big value of the radius $r$ at the joints increases the effective shear modulus, but $r=0.420$ mm is the limit of the machining method we are using.
\begin{table}[ht]\caption{Parameters of the unit cells corresponding to different values of refractive index as shown in Fig.\ \ref{layered}.} \label{tab}
\begin{ruledtabular}
\begin{tabular}{c c c c c c}
  $n_{\text{eff}}$ & $l$ (mm) & $t$ (mm) & $a$ (mm) & $q$ (mm) & $r$ (mm)  \\  
  \hline
  1.000 & 9.708 & 0.693 & 6.025 & 2.184 & 0.420  \\ 
  0.977 & 9.708 & 0.708 & 5.844 & 2.184 & 0.420  \\ 
  0.910 & 9.708 & 0.761 & 5.295 & 2.184 & 0.420  \\
  0.810 & 9.708 & 0.851 & 4.451 & 2.184 & 0.420  \\
  0.690 & 9.708 & 0.994 & 3.397 & 2.184 & 0.420 \\
  0.561 & 9.708 & 1.213 & 2.177 & 2.184 & 0.420  \\
\end{tabular}
\end{ruledtabular}
\end{table}

\begin{figure}[ht]
\centering
     \includegraphics[width=\columnwidth]{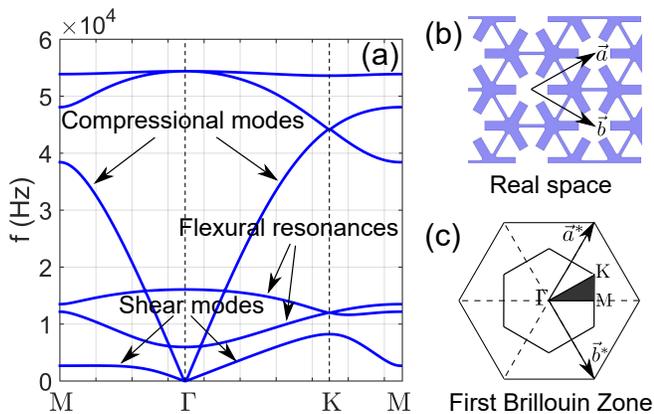} 
\caption{Band diagram (a) of the unit cell (b) at the center ($n_{\text {eff}}=1$) along the $\Gamma-M-K$ of the first Brillouin zone (c). }\label{band}
\end{figure}

The GRIN lens is comprised of the six types of unit cells, the minimum cutoff frequency is limited by the unit cell with thinnest plates, i.e. $n_{\text {eff}}=1$, therefore it is essential to examine its band structure. The band diagram as shown in Fig.\ \ref{band} is calculated using Bloch-Floquet analysis in COMSOL. 
The directional band gap along the incident direction occurs near $40$ kHz, this sets the upper limit of the lens. The lens is designed following an index gradient, therefore the low frequency focusing capability is limited due to the high frequency approximation nature of the ray theory. Although bending modes exist at low frequency range, they do not cause much scattering due to sufficient shear modulus which prevents the structure from flexure. \cite{Cai2016} We expect the lens to be capable of focusing underwater sound over a broadband from $10$ kHz to $40$ kHz.

\section{\label{results}Simulation results}
\begin{figure*}[ht]
\centering
     \includegraphics[width=0.8\textwidth]{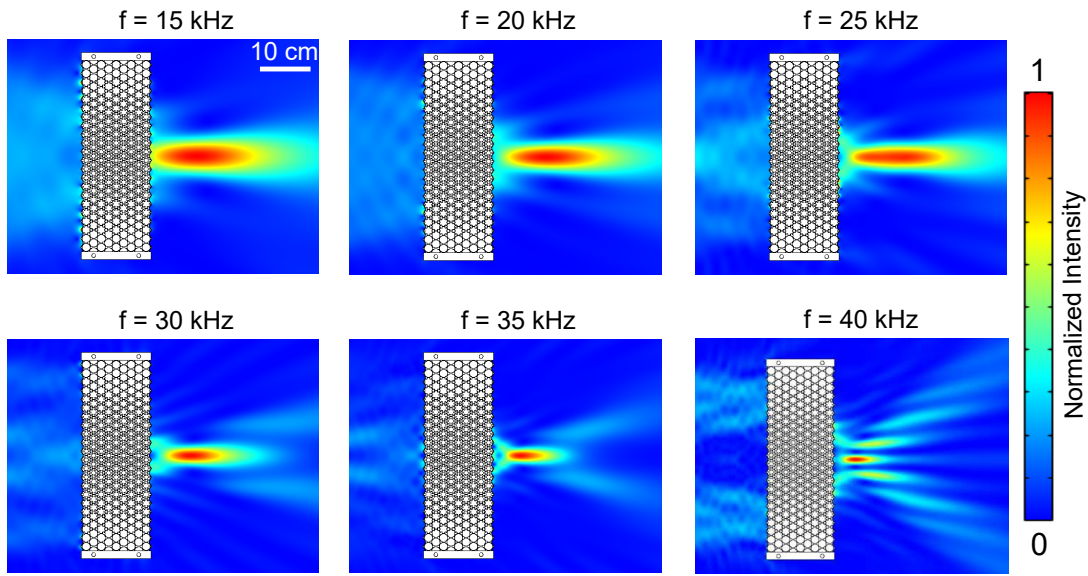} 
\caption{Simulation results for a plane wave normally incident from the left side. Each plot shows the normalized steady state intensity at $15$ kHz, $20$ kHz, $25$ kHz, $30$ kHz and $35$ kHz and $40$ kHz, respectively.}\label{intensity}
\end{figure*}

The lens is formed by combining all the designed unit cells together following the reduced aberration profile. The length of the lens is $40$ cm, and the width is $13.7$ cm. The material of the lens is aluminum as we described in the previous section. The GRIN is permeated with air and immersed in water so that only structural wave is allowed in the lens. Full wave simulations were done to demonstrate the broadband focusing effect using COMSOL Multiphysics. Figure \ref{intensity} shows the intensity magnitude normalized to the maximum value at the focal point from $15$ to $40$ kHz.
A Gaussian beam is normally incident from the left side, and the focal point lies on the right side of the lens. It is clear that the lens works over a broad range of frequency. In the focal plane, the high intensity focusing region moves towards the lens as the frequency increases. This is not surprising as we explain as follows. The low frequency focusing capability is limited due to the high frequency approximation nature of the index gradient, while the high frequency is limited because the longitudinal mode becomes dispersive as shown in Fig.\ \ref{band}, i.e. the effective speed is reduced. The best operation frequency of the lens is found to be near $20$ kHz where the longitudinal mode is non-dispersive. The cutoff frequency is near $40$ kHz as predicted in the band diagram. 

\begin{figure}[ht]
\centering
     \includegraphics[width=0.9\columnwidth]{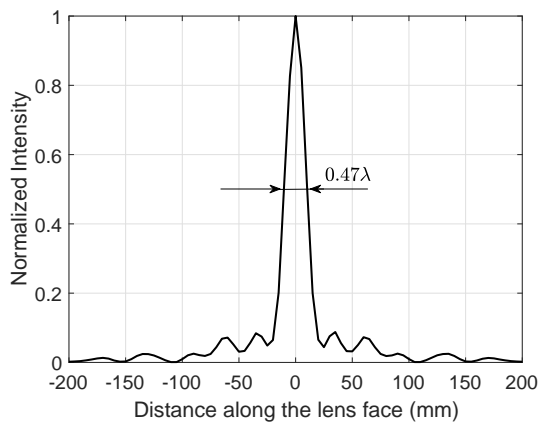} 
\caption{Focusing capability at $35$ kHz. The plot shows the simulated normalized intensity along the direction parallel to the lens face and through the focal point.}\label{lineplot}
\end{figure}

The as-designed lens has minimized side lobes comparing to conventional diffractive lens. Diffractive acoustic lenses are usually designed by tuning the impedance of each channel to achieve certain phase delay. However, the transmitted amplitudes are different so that it is hard to cancel out the side lobes caused by aperture diffraction. The main advantage of the GRIN lens is that it redirects the ray paths inside the lens, and reduces the diffraction aperture to a minimal size at the exiting face of the lens. Figure \ref{lineplot} shows the normalized intensity magnitude across the focal point along the lens face.
The width of the intensity profile at half of its maximum is only $0.47\lambda$ at $35$ kHz. The focal distance at this frequency is about $5$ cm. It is also clear that the intensity magnitudes of the side lobes are all below $1/10$ of the maximum value so that our GRIN lens is nearly side lobe free. 

\begin{figure}[ht]
\centering
     \includegraphics[width=1\columnwidth]{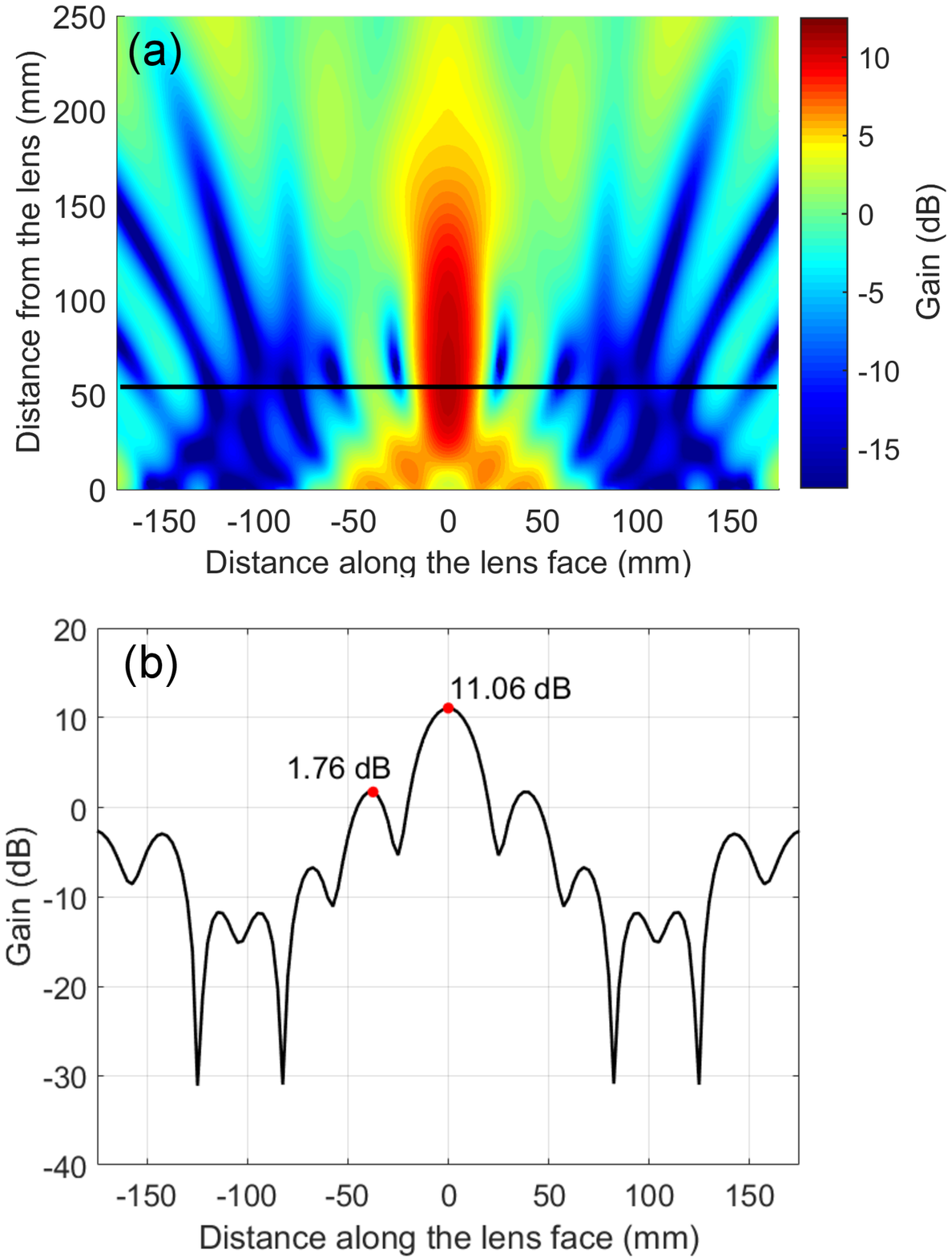}
\caption{Simulated sound pressure level gain (dB) at $33.5$ kHz. The gain in the focal plane is shown in (a), the gain through the focal point along the horizontal line is shown in (b). }\label{gain335}
\end{figure}

As we mentioned in Sec.\ \ref{Homo}, the as-designed pentamode GRIN lens is impedance matched to water so that it is acoustically transparent (back-scattering free) to a normally incident plane wave. This feature should result in a very high gain at the focal plane. Figure \ref{gain335} shows the simulated sound pressure level (SPL) gain at $33.5$ kHz over the focal plane. 
This plot is generated by subtracting the simulated SPL without the lens from the SPL with the lens for normally incident plane wave beams. It is remarkable that the maximum gain at $33.5$ kHz is as high as $11.06$ dB which is hard to achieve for a diffractive lens, especially for a 2D device. The advantage of the pentamode GRIN is that it can achieve high gain and minimal side lobes at the same time, however, minimizing the side lobes for a diffractive lens is usually at the cost of introducing high impedance mismatch.

\begin{figure}[ht]
\centering
     \includegraphics[width=\columnwidth]{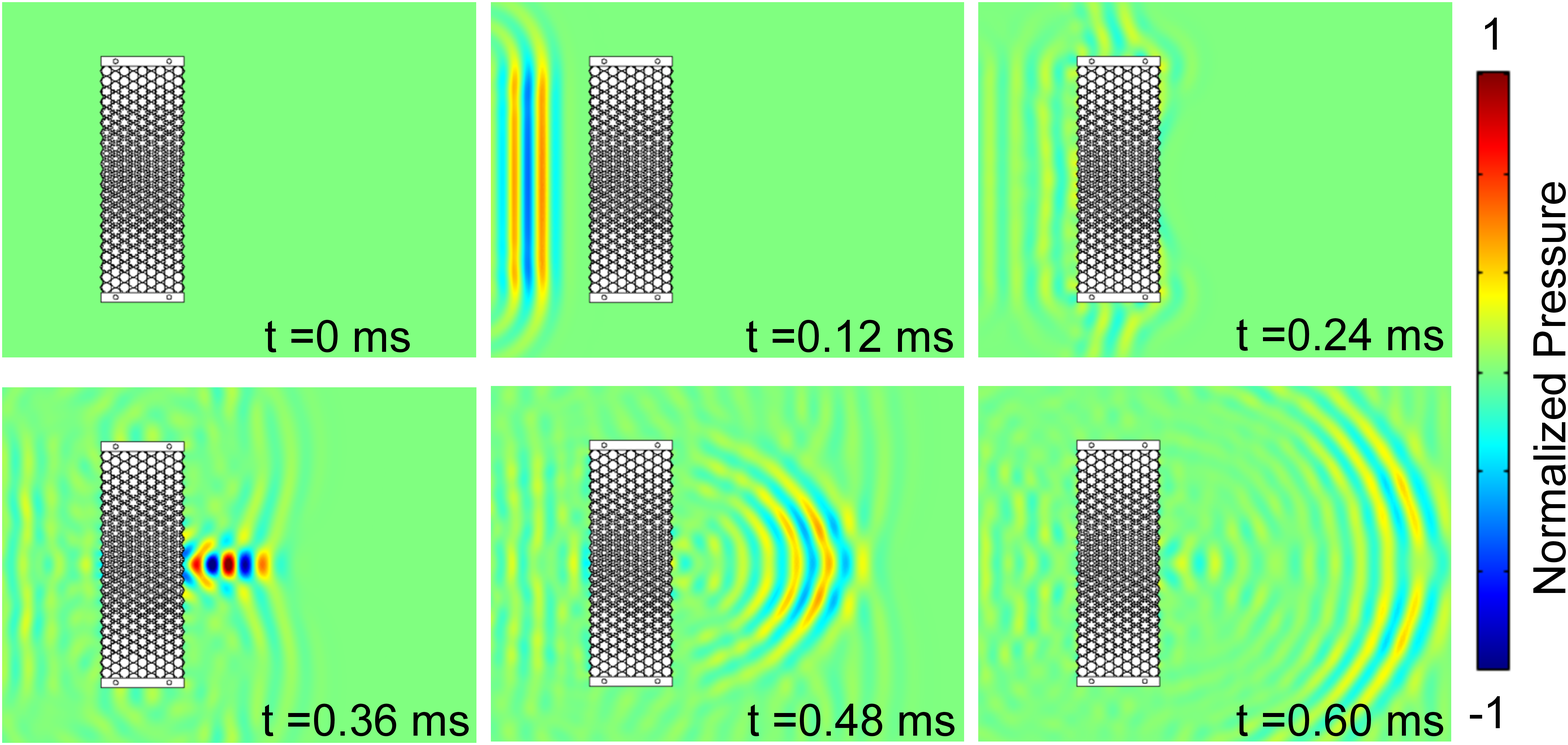} 
\caption{Simulated transient pressure wave propagation at $30$ kHz. The six figures correspond to times $0$ ms, $0.12$ ms, $0.24$ ms, $0.36$ ms, $0.48$ ms and $0.60$ ms, respectively. The pressure is normalized to the maximum at $t=0.36$ ms. }\label{timed}
\end{figure}

Unlike the diffractive metasurfaces, which only work at the steady state, the pentamode GRIN lens is also capable of focusing a plane wave pulse. Figure \ref{timed} shows the simulated pressure variations at each time frame. The acoustic pressure in all the six plots are normalized to the maximum at $t=0.36$ ms.
Two cycles of a plane wave pulse are incident from the left side at the central frequency of $30$ kHz. The wave moves towards the lens and then transmits through the lens as shown in each time frame. The wave focuses on the right side of the lens and starts to spread out when $t=0.36$ ms. It is also easy to see from the third plot, i.e. $t=0.24$ ms, that the reflection from the water-lens interface is almost negligible. 

\section{\label{experi}Experiments}
\subsection{Experimental apparatus}
\begin{figure}
	\includegraphics[width=0.95\columnwidth]{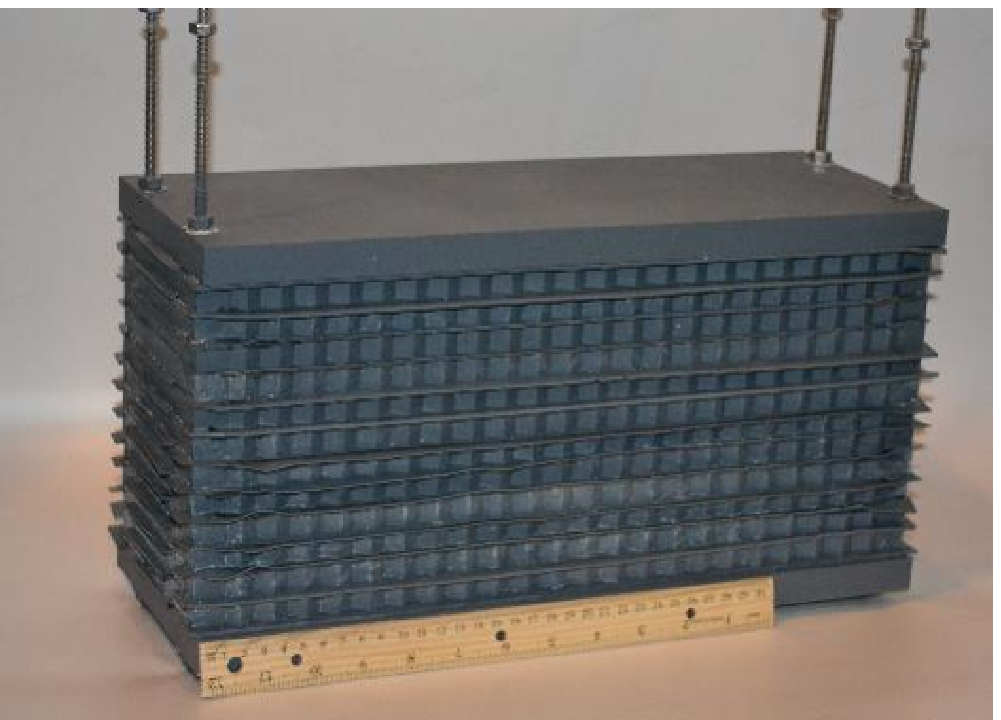}
	\caption{Photograph of assembled lens showing the two 2 cm thick aluminum end caps with four steel rods compressing the twelve lens pieces and alternating 1 mm thick neoprene gaskets. A 12 in ruler is included for scale.}
	\label{fig:assembledlens}
\end{figure}

The GRIN lens pictured in Fig.~\ref{fig:assembledlens} was fabricated using an abrasive water jet cutting twelve pieces $1.5$ cm-thick aluminum plates. 
The dimensions of the plates were measured and compared to the specified dimensions in Table \ref{tab}. The maximum discrepancy was $0.5$ mm from the desired dimension with an average difference of $0.2$ mm. These deviations were noted as a source of possible error in the experimental data. The as-tested lens is constructed by assembling twelve fabricated plates so that the inside could be air-tight. Rubber gaskets were cut out of neoprene sheets to provide a $1$ cm rubber border around the perimeter of each lens piece and the outer edge of the top and bottom of each piece was lined with a layer of electrical tape and double sided tape to hold the gaskets in place. The layers were then placed on top of one another alternating with rubber gaskets. Two blocks of aluminum measuring $40.0$ cm by $15.25$ cm, and $2$ cm thick were placed on the top and bottom of the stacked pieces and were compressed together using nuts and washers with four steel rods. The compression of the gaskets provided a means of overcoming the surface irregularities on the perimeters of each piece to prevent leakage.

All the experimental measurements were done in a rectangular indoor tank approximately $4.5$ m in depth with a capacity of $459$ m$^3$ surrounded by cement walls with a sand covered floor. The tank is filled with fresh water and the temperature is assumed to be of negligible variance between tests. An aluminum and steel structure was constructed to secure the lens and source separated by $1$ cm at a centerline depth of $68.5$ cm. The structure was attached to a hydraulically actuated cylinder which held the components at a consistent desired depth for the duration of testing. An exponential chirp at 1 ms in duration with a frequency range of $10$ kHz to $70$ kHz was used as the excitation signal and the signal was repeated every $100$ ms. 

\begin{figure}
	\includegraphics[width=0.95\columnwidth]{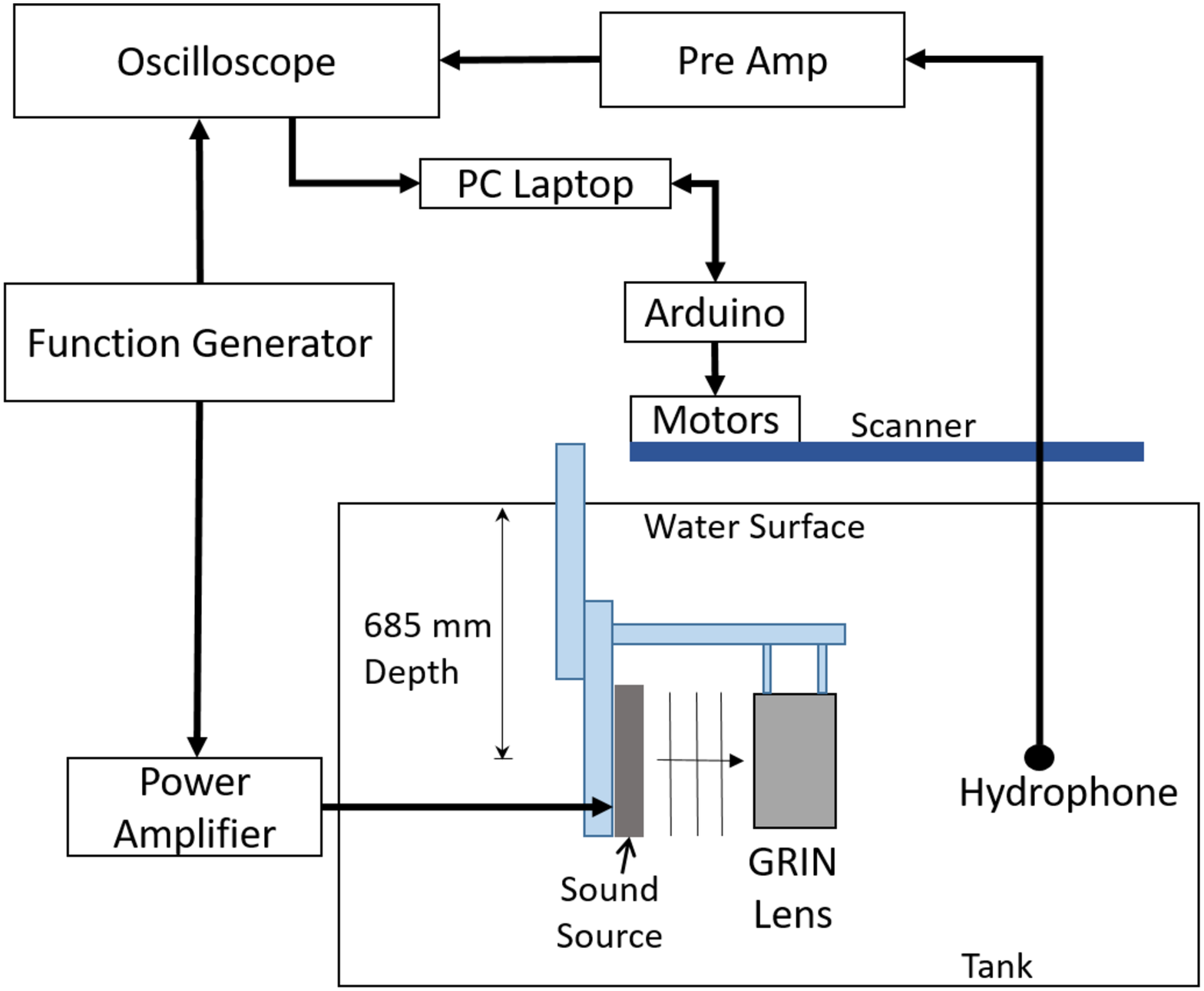}
	\caption{Schematic of the experimental test apparatus is shown. The aluminum and steel structure supports the source and lens with a separation distance of approximately 1 cm. A hydraulic column holds the structure at a constant depth of 0.685 m referenced to the vertical centerline of the source and lens. The distance between the sound source and the GRIN Lens is exaggerated in the figure.}
	\label{fig:apparatus}
\end{figure}

\begin{figure}
	\includegraphics[width=0.95\columnwidth]{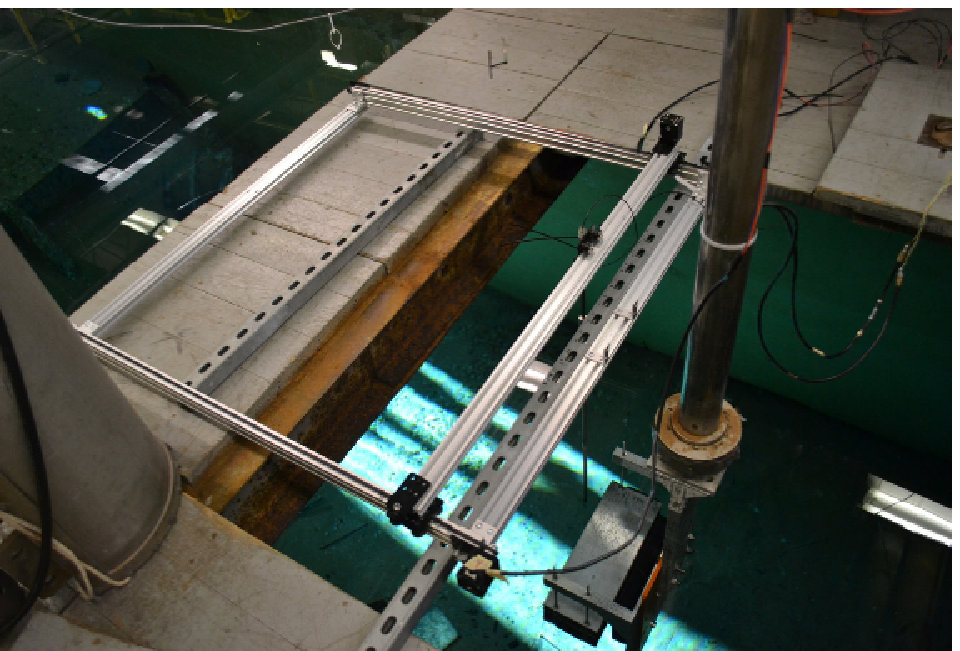}
	\caption{Full test apparatus shown during experimentation. The extruded aluminum framework is the Arduino controlled hydrophone scanner. On the right side of the figure is the GRIN lens supported by the suspension structure at depth via the hydraulic column.}
	\label{fig:realapp}
\end{figure}

An automated scanning process as shown in Fig.~\ref{fig:apparatus} was used to acquire hydrophone amplitude measurements.
Three stepper motors controlled by MATLAB via an Arduino Uno moved a rod with a RESON TC4013 Hydrophone attached to the end through a rectangular area in front of the GRIN lens. The scan area was collinear with center-line plane of the source and GRIN lens at a depth of 685 mm. Figure \ref{fig:realapp} shows the experimental apparatus, including the support structure, GRIN Lens, and the planar hydrophone scanner.
The area was 31.0 cm parallel to the lens face by 20.0 cm perpendicular to the lens face. The step size was set to 5 mm which resulted in 2,583 data points.
As the hydrophone moved to each location, a pause of 2 seconds was initiated by the MATLAB program to negate rod dynamics due to the swaying caused by the scanner motion in the water. Voltage outputs were acquired from the oscilloscope and stored in an excel spreadsheet labeled for its exact location in the scan area. After each point had voltage data, the scanning program terminated after approximated 4.5 hours of run time. This process was completed with both the lens and the source, and another case with just the source. This would allow the effects due to the inclusion of the lens to be quantified by comparing the amplitude changes between the source only case and the source-lens case.

\begin{figure}
    \centering
	\includegraphics[width=1\columnwidth]{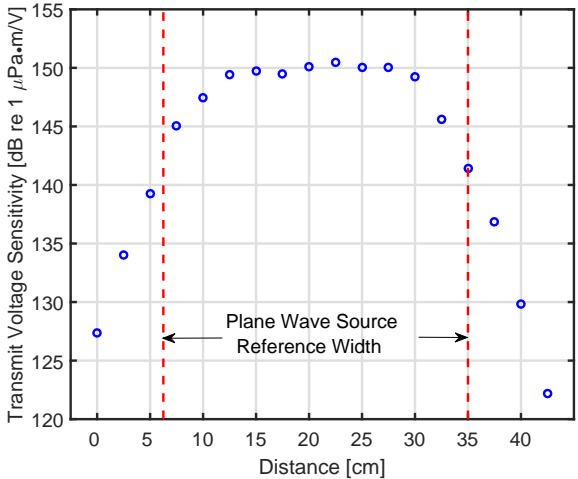}
	\caption{Planarity verification of the source. TVR was determined at evenly spaced locations at a constant distance of 9.5 cm from the source face. Source reference width is noted by the vertical dotted lines.}
	\label{fig:tvrsource}
\end{figure}

To begin simulation verification, a source capable of generating constant amplitude acoustic waves was constructed and tested. The source is 29.5 cm in width, 22.9 cm in height, and 6.4 cm in depth. The planarity was verified by submerging the source at a depth of 68.5 cm measured from centerline and measuring pressure amplitude using an omni-directional hydrophone. The test signal was prescribed to be a sinusoidal pulse at a frequency of 35 kHz and amplitude of 2 Volts peak-to-peak for 15 cycles continuously repeating every 100 ms.  The Hilbert transform was taken of the hydrophone measurement and the mean amplitude of the Hilbert transform was calculated for the steady state region of the signal. The transmit voltage response (TVR) of a transducer is the amount of sound pressure produced per volt applied and is calculated using 
\begin{equation}
\label{eqn:TVR}
\textit{TVR} =
20\textrm{log}_{10}\left(\displaystyle\frac{V_{\textrm{out}}R_{\textrm{meas}}}{V_{\textrm{in}}R_{\textrm{ref}}}\right) - \textit{RVS}_{\textrm{cal}},
\end{equation}	
where $V_{\textrm{out}}$ is the output voltage from the hydrophone, $V_{\textrm{in}}$ is the voltage applied to the transducer, $R_{\textrm{meas}}$ is the separation distance between the transducer and the hydrophone, $R_{\textrm{ref}}$ is the reference distance set to 1 m, and $RVS_{\textrm{cal}}$ is receive sensitivity of the calibrated hydrophone taken from the hydrophone documentation. The $R_{\textrm{meas}}$ distance was set to 9.5 cm, $V_{\textrm{in}}$ was 2 Vpp, and $RVS_{\textrm{cal}}$ was 211 dB/$\mu$Pa. The planarity amplitude test results are shown in  Fig. \ref{fig:tvrsource}.
The amplitude measurements show that there is relatively consistent planarity across the aperture of the source face. However, as the boundaries of the source are reached, the amplitude reduces by approximately 7 dB. Even though the amplitude decreases, the source operates effectively enough to be used to verify the GRIN lens simulations. It should be noted that source planarity may be a cause for a reduction in amplitude shown in the GRIN lens experiment because the width of the lens extends outside the borders of the source width.

\subsection{Data Processing}\label{DP}

\begin{figure}
	\includegraphics[width=1\columnwidth]{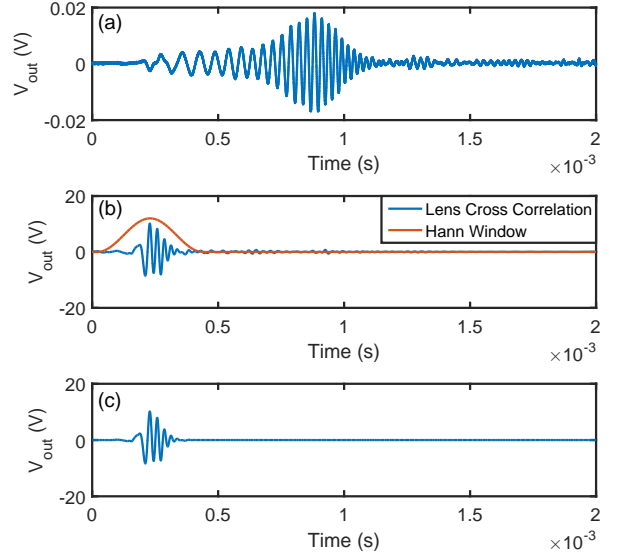}
	\caption{The raw hydrophone voltage signal is shown in (a).  The hydrophone signal cross correlated with the input signal is shown in (b) with the blue curve.  Applying the red window function results in the signal shown in (c). The window function amplitude has been exaggerated to better pictorially represent where it is applied.}
	\label{fig:procex}
\end{figure}

For both the source-only case and the source-lens case, the cross-correlation between the input signal and the voltage output from the hydrophone was determined. A Hann window was applied to the cross-correlation over the direct path form the source. This removed any reflections from the water surface of the tank or diffraction from the source interaction with the edges of the lens from contaminating the results. An example of this process is shown in Fig. \ref{fig:procex}.
The Fourier transform of the cross-correlation for both cases was then found. The gain was then calculated by means of Eq.\ \ref{eqn:gain},
\begin{equation}
\label{eqn:gain}
\textit{G} =
20\textrm{log}_{10}\left(\displaystyle\frac{X_{\textrm{lenswin}}}{X_{\textrm{sourcewin}}}\right)
\end{equation}	
where $G$ is the gain at a particular scan point and frequency, $X_{\textrm{lenswin}}$ is the windowed cross-correlation from the source-lens case, and $X_{\textrm{sourcewin}}$ is the windowed cross-correlation from the source-only case.

\subsection{Measurement results}
\begin{figure*}
\includegraphics[width=0.95\textwidth]{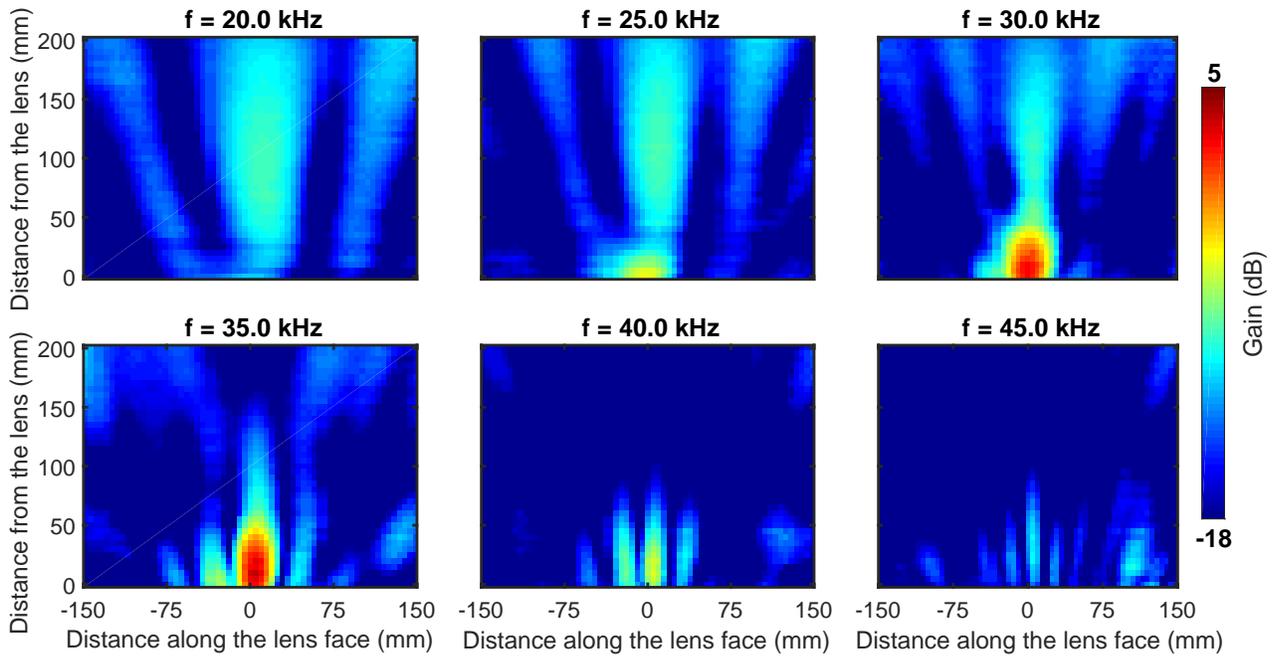}
\caption{These are plots of the gain (dB) exhibited due to the inclusion of the GRIN Lens. The plots are oriented from a top-view orientation of the entire scan area. Each plot is a single frequency, and frequencies increase by 5 kHz starting with 20 kHz in the top left and ending with 45 kHz in the bottom right. The plots have been rotated $90^{\circ}$ in the counter-clockwise direction from the orientation in Fig.\ \ref{intensity}.}
\label{fig:gainplots}
\end{figure*}

\begin{figure}
	\includegraphics[width=0.95\columnwidth]{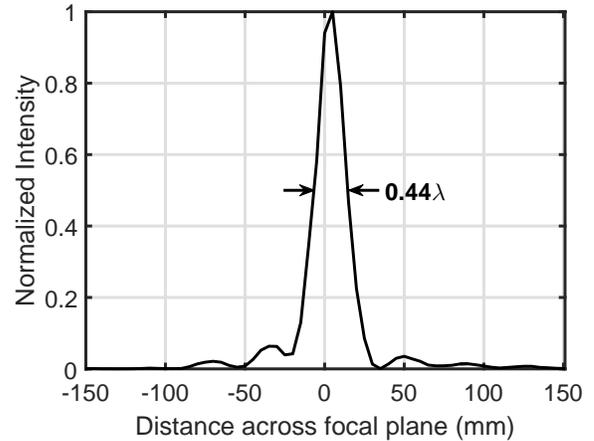}
	\caption{Normalized intensity through the focal plane at 35 kHz. Beamwidth at 0.5 of the normalized intensity was calculated to be 0.44$\lambda$ as noted in the figure.}
	\label{fig:focalplane35k}
\end{figure}

As outlined in Sec.~\ref{DP}, the gain was measured by finding the amplitude difference between the source-only and the source-lens cases. The measurements at frequencies from 20 to 45 kHz are shown in Fig.~\ref{fig:gainplots}.
The amplitude scale represents the gain at each hydrophone location in decibels. The general shape of the beam pattern shows a clear focusing tendency of the lens, especially in the 30-40 kHz range.The data shows evidence of a focused beam pattern forming at 20 kHz with approximately -5 dB of gain at the focus. As the frequency increases, the beam becomes narrower and the gain increases to peak levels at 30 and 35 kHz. There is also evidence a stop band is approached as the frequency approaches 45 kHz.
Figure \ref{fig:focalplane35k} shows the beam pattern of the normalized intensity through the focus for 35 kHz. Significant side lobe amplitude reduction is evident, and the beam width is 0.44$\lambda$ with the speed of sound in fresh water assumed to be 1480 m/s.

\begin{figure}
\includegraphics[width=0.95\columnwidth]{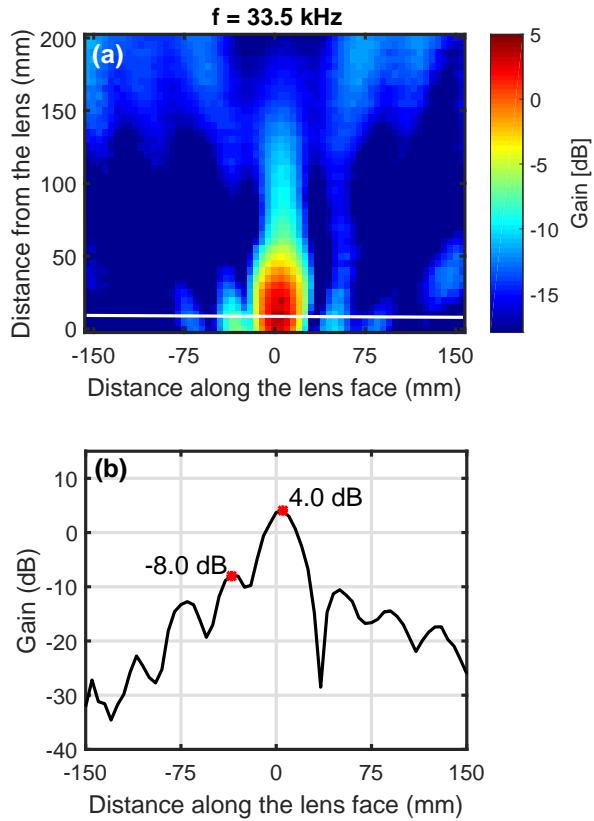}
	\caption{Experimental sound pressure level gain (dB) at $33.5$ kHz. The gain in the focal plane is shown in (a), the gain through the focal point along the horizontal line is shown in (b).}
	\label{fig:maxgain}
\end{figure}

The maximum gain through the frequency range was determined to be at 33.5 kHz as shown in Fig.~\ref{fig:maxgain}.
To better quantify the data, a cross section of the amplitude data was extracted from upper plot in Fig.~\ref{fig:maxgain} for a constant distance from the lens through the peak gain of focus. The maximum gain was observed to be 4.0 dB and the beam pattern was found to have 12 dB of sidelobe amplitude reduction compared to the focus as shown in the lower plot in Fig.~\ref{fig:maxgain}.

The as-designed and as-tested lenses both work over a broad range of frequency. Figures \ref{intensity} and \ref{fig:gainplots} both show that the focal point moves toward the lens with the increase of frequency as predicted from the band diagram. It is also clear that the side lobe suppression ability of the GRIN lens in both simulation and experiment agree to a remarkable degree as can be seen from Figs.~\ref{lineplot} and \ref{fig:focalplane35k}, where the magnitude of the intensity of the side lobes are all lower than 1/10 of the maximum magnitude at the focal point. It is noted that the power magnification at the focal point have certain differences between simulations and experiments. These discrepancies are mainly due to the fabrication of the lens as we explain in the following section.

\subsection{Sources of Error and Discussion}
Potential error in the experiment was noted as data was taken. First, the source itself had acceptable planarity, but as shown in Fig.~\ref{fig:tvrsource}, there is amplitude reduction at the edges of the source. This results in the outside portions of the lens to have less contribution to the focusing beam pattern than was assumed in the simulation. The lens pieces themselves have a machining tolerance that also affects the mass and stiffness properties of the architecture. With an effectively random distribution of tolerances throughout the assembled lens, the altered effective index distribution may cause some variability in the focal distance. 

During the scanning process, the hydrophone rod moved from location to location to acquire data. In order to protect the scanning components, the scanner could not be submerged underwater, but the depth of the lens and source were desired to be at the greatest depth possible to eliminate contamination by reflections from the water surface. However, this resulted in the hydrophone rod to have a length longer than the depth of the lens with a single attachment point at its extreme. As the location changed, the resistance of the water caused the lens to sway momentarily during the beginning of each measurement potentially affecting the results.

The lens construction also includes the rubber gaskets between each piece. Some excess rubber was necessary to extend over the perimeters of each lens piece to ensure a watertight seal. However, this excess rubber results in an impedance mismatch between the lens face and the surrounding water. This causes a reflection of wave energy at both the front and back faces of the lens and inevitably causes a reduction of energy that should reach the focus. The surface impedance mismatch induced by the alternating layers causes a lower gain than expected. Moreover, the impedance mismatch could cause focal distance shift even though the index distribution still follows the modified profile as we described in the introduction.

These sources of error support the observed differences between the simulation and experiment with the most noticeable being the lower gain obtained via the experiment. There is a 5 dB deficit from the simulations and can be attributed to the excess rubber causing and impedance mismatch with high confidence. 
\section{\label{Conc}Conclusion}

In conclusion, we have designed and fabricated a pentamode GRIN lens based on a modified secant index profile. We have experimentally demonstrated its  broadband focusing effect for underwater sound. The unit cells are tuned to be  impedance-matched  to water so that the GRIN lens is capable of focusing sound with minimized aberration. Moreover, the physics behind the GRIN lens makes it possible to focus sound at both steady state and transient domain. The mismatch of the focal distance in simulation and experiments is due to the  accuracy of the waterjet machining process and the assembly method which altered the refractive index. This issue could  be successfully resolved by using more advanced fabrication methods such as wire EDM or 3D metal printing. The design method can also be easily extended to the design of anisotropic metamaterials such as directional screens and acoustic cloaks.

\section*{Acknowledgments}

This work was supported by ONR through MURI Grant No.\ N00014-13-1-0631. 



\end{document}